\documentclass{PoS}
\usepackage{graphicx}
\usepackage{subfigure}

\title{The Landau gauge gluon propagator at zero and finite temperature: accounting for the combined finite lattice spacing and finite volume effects}

\ShortTitle{The Landau gauge gluon propagator at zero and finite temperature}

\author{O. Oliveira\\
        CFC, Departamento de F\'{\i}sica, Universidade de Coimbra, 3004-516 Coimbra, Portugal\\
        E-mail: \email{orlando@teor.fis.uc.pt}}

\author{\speaker{P. J. Silva}\\
        CFC, Departamento de F\'{\i}sica, Universidade de Coimbra, 3004-516 Coimbra, Portugal\\
        E-mail: \email{psilva@teor.fis.uc.pt}}

\abstract{In the past years a good comprehension of the infrared gluon propagator has been achieved, with a good qualitative agreement between lattice results and Dyson-Schwinger equations. However, lattice simulations have been performed at physical volumes which are close to 20 fm but using a large lattice spacing. The interplay between volume effects and lattice spacing effects has not been investigated. Here we aim to fill this gap and address how the two effects change the gluon propagator in the infrared region. Furthermore, we provide infinite volume extrapolations which take into account the finite volume and finite lattice spacing. 

We also report on preliminary results for the gluon propagator at finite temperature. }

\FullConference{The 30th International Symposium on Lattice Field Theory\\
		 June 24 -- 29,  2012\\
		 Cairns, Australia}

\begin{document}

\section{Introduction and motivation}

In  a Quantum Field Theory, the knowlegde of all Green's functions allows a complete description of the theory. In QCD, propagators of the fundamental fields encode information about non-perturbative phenomena, like confinement and dynamical chiral symmetry breaking.

Here we study the gluon propagator in Landau gauge, at both zero and finite temperature, using lattice simulations. 

\section{The gluon propagator at zero temperature}

In this section we study the gluon propagator using lattice simulations for various lattice volumes and lattice spacings. Although there is some support for D(0)=0 \cite{ratios, bounds}, recent large volume lattice simulations, close to 20 fm, have claimed a finite non-vanishing gluon propagator at zero momentum \cite{brasil, BMA09}. In order to be able to simulate such large volumes, the reported simulations used a large lattice spacing $\sim 0.2$ fm. Since there is no systematic study about the effect of such a large lattice spacing in the propagator, our first goal is to investigate the interplay between volume effects and lattice spacing effects. Furthermore, we also consider the extrapolation of the propagator to the infinite volume limit. 

\subsection{Lattice effects}

\begin{figure}[t]
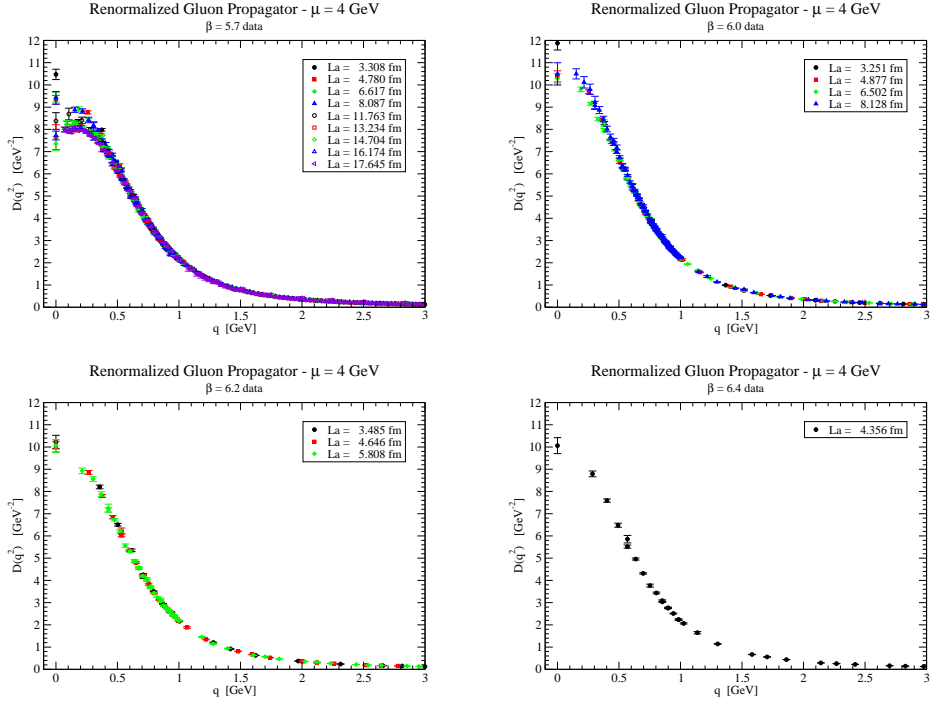
 %  figure placement: here, top, bottom, or page
   \centering
   \subfigure{ \includegraphics[scale=0.235]{figures/glue_R4GeV_B5.7.eps} } \qquad
   \subfigure{ \includegraphics[scale=0.235]{figures/glue_R4GeV_B6.0.eps} }

   \subfigure{ \includegraphics[scale=0.235]{figures/glue_R4GeV_B6.2.eps} } \qquad
   \subfigure{ \includegraphics[scale=0.235]{figures/glue_R4GeV_B6.4.eps} }
  \caption{Renormalized gluon propagator for $\mu = 4$ GeV for all lattice simulations.}
   \label{fig:gluevol}
\end{figure}

In figure \ref{fig:gluevol}, we show our results for the gluon propagator, renormalized at $\mu=4$GeV --- see \cite{OliSi12} for details and the lattice setup\footnote{Simulations in this section have been performed with MILC code \cite{milc}.}. For comparison purposes, we also show data obtained by the Berlin-Moscow-Adelaide collaboration \cite{BMA09}. Note that data in the same plot has the same lattice spacing, while varying the lattice volume. The plots show that, in the infrared region, the gluon propagator decreases as the lattice volume increases. 

\begin{figure}[t]
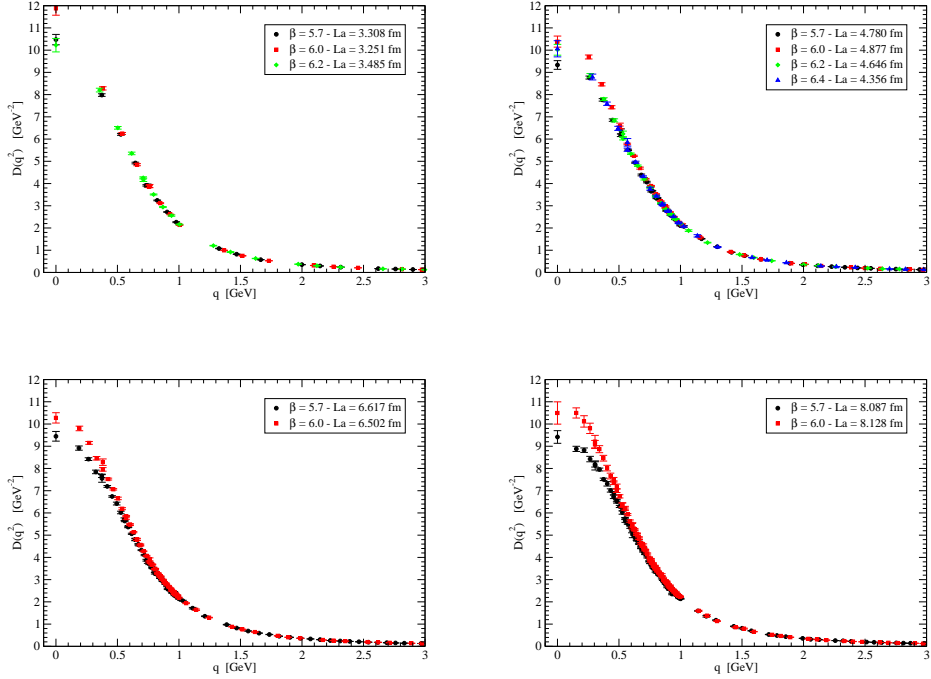
 %  figure placement: here, top, bottom, or page
   \centering
   \subfigure{ \includegraphics[scale=0.235]{figures/glue_R4GeV_V3.3fm.eps} } \qquad
   \subfigure{ \includegraphics[scale=0.235]{figures/glue_R4GeV_V4.6fm.eps} }

\vspace{0.6cm}
   \subfigure{ \includegraphics[scale=0.235]{figures/glue_R4GeV_V6.6fm.eps} } \qquad
   \subfigure{ \includegraphics[scale=0.235]{figures/glue_R4GeV_V8.1fm.eps} }
  \caption{Comparing the renormalized gluon propagator at $\mu = 4$ GeV for various lattice spacings and similar physical volumes.}
   \label{fig:gluespac}
\end{figure}

Whereas in figure \ref{fig:gluevol} we compare data with the same lattice spacing, in figure \ref{fig:gluespac} we plot data with similar physical volumes. This allow us to study how the propagator changes with the different lattice spacings, keeping a constant physical volume. We have considered 4 different volumes, namely $\sim$3.3,$\sim$4.6,$\sim$6.6, and $\sim$8.1 fm. 

The first thing to note in figure \ref{fig:gluespac} is that, for momenta above $\sim$900 MeV, the propagator is well-defined, in the sense that the data define a unique curve. We can therefore claim that the renormalization procedure has been able to remove lattice artifacts for the high momenta region.

Furthermore, figure \ref{fig:gluespac} shows that the simulations performed
 with smaller $\beta$ values (i.e. larger lattice spacings) underestimate 
the gluon propagator in the infrared region. Indeed, by comparing figures 
\ref{fig:gluevol} and \ref{fig:gluespac} one can conclude that the corrections 
due to the finite lattice spacing seem to be larger than the corrections due 
to the finite volume.  Moreover, from the above observations one can claim that the data from the Berlin-Moscow-Adelaide collaboration provide a lower bound for the gluon propagator in the continuum.
 
\subsection{Zero momentum gluon propagator}

\begin{figure}
\begin{center}
\includegraphics[width=0.7\textwidth]{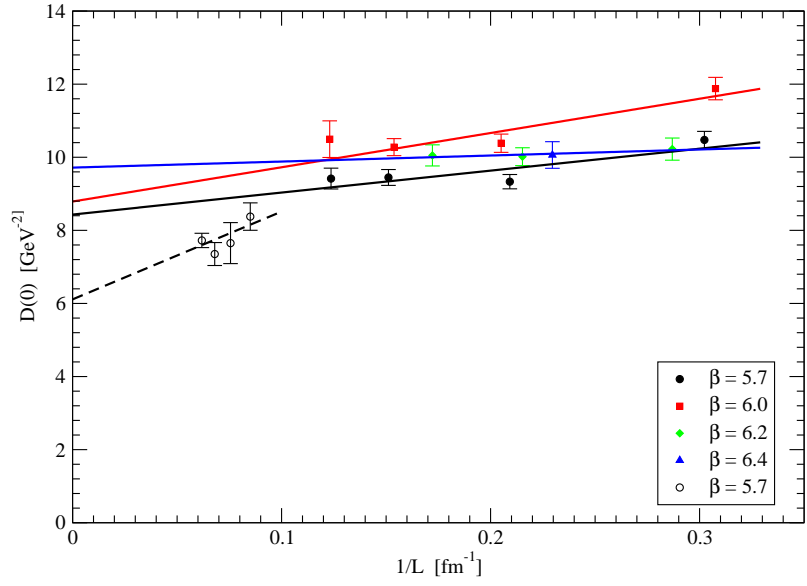}
\end{center}
\caption{Linear extrapolation of $D(0)$ to the infinite volume.}
\label{Dzero}
\end{figure}

Figure \ref{Dzero} shows the zero momentum gluon propagator as a function of $1/L$. For each set of $D(0)$ values with the same $\beta$, one can consider the extrapolation to the infinite volume. Here we consider the ansatz 

\begin{equation}
D(0)=\frac{c}{L}+ D_{\infty}(0)
\end{equation}

The results for $D_{\infty}(0)$ are $8.43\pm0.61 \mathrm{GeV}^{-2}$ ($\chi^2/d.o.f.=2.6$) for $\beta=5.7$, $8.79\pm0.64 \mathrm{GeV}^{-2}$ ($\chi^2/d.o.f.=1.7$) for $\beta=6.0$, and $9.72\pm0.25 \mathrm{GeV}^{-2}$ ($\chi^2/d.o.f.=0.1$) for $\beta=6.2$. In what concerns the data from the Berlin-Adelaide-Moscow collaboration, the fit provides $D_{\infty}(0)=6.1\pm1.4 \mathrm{GeV}^{-2}$ ($\chi^2/d.o.f.=1.3$). Note that this linear extrapolation does not provide a coherent picture of all the data sets. However, one can claim a $D_{\infty}(0)$ in the range 6--10 $\mathrm{GeV}^{-2}$. 
Note, however, that the results reported in \cite{ratios, bounds} do not exclude completely a vanishing gluon propagator at zero momentum.

\subsection{Extrapolation to the infinite volume limit}

Having in mind the extrapolation of our results to the infinite volume limit, we consider fits of the lattice data to the functional form

\begin{equation}
D(q^2)=Z\frac{q^2+M_1^2}{q^4+M_2^2q^2+M_3^4}.
\label{extrafit}
\end{equation}

\begin{figure}
\begin{center}
\includegraphics[width=0.7\textwidth]{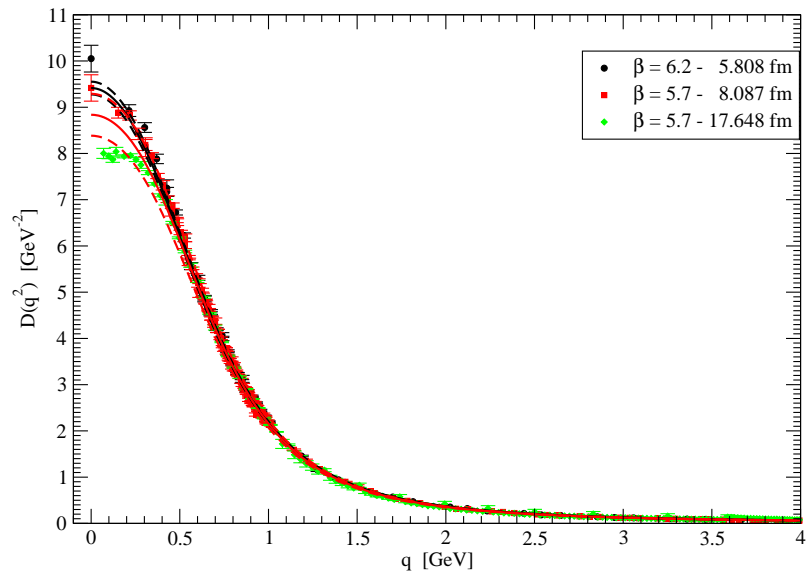}
\end{center}
\caption{Extrapolated propagators to the infinite volume.}
\label{figext}
\end{figure}

Note that for Z=1, the above expression is the tree level prediction of the so-called refined Gribov-Zwanziger action and, as shown in \cite{lusobelga}, it describes the infrared lattice gluon propagator up to momenta $\sim1.5\textrm{GeV}$. The extra parameter Z allows to fit the lattice data from 0 up to 4GeV --- see \cite{OliSi12} for details.  Then we combine all volumes for a given $\beta$ and perform a linear extrapolation in $1/L$ to the infinite volume of each parameter independently. The extrapolated parameters can be seen in table \ref{extrapol}. All extrapolations have a $\chi^2/d.o.f.$ below 1.25 with the exception of $M_3^4$ for $\beta=6.0$; for this reason, no information is given about this parameter in table \ref{extrapol} for the $\beta=6.0$ case.  In figure \ref{figext}, we show the extrapolated propagators and compare with the largest lattice volume available in each case.

\begin{table}
\begin{center}
\begin{tabular}{cccccc}
\hline
 $\beta$ & $Z$       &  $M_1^2$ &  $M_2^2$ &  $M_3^4$ &  $D(0)$  \\
\hline
5.7      & 0.821(10) & 4.09(17) & 0.558(36) & 0.380(11) & 8.84(45) \\
6.0      & 0.830(13) & 4.01(16) & 0.565(46) &  ---      &  --- \\
6.2      & 0.83333(17) & 4.473(21) & 0.704(29) & 0.3959(54) & 9.42(14) \\
\hline
\end{tabular}
\end{center}
\caption{Extrapolation to the infinite volume limit. $D(0)$ is given in $\textrm{GeV}^{-2}$.}
\label{extrapol}
\end{table}

\section{The gluon propagator at finite temperature}

In this section, we focus on the calculation of the gluon propagator at finite temperature. On the lattice, finite temperature is introduced by reducing the temporal extent of the lattice, i.e. we work with lattices $L_s^3 \times L_t$, with $L_t \ll L_s$. The temperature is defined by $T=1/a L_t$.

At finite temperature, the Landau gauge gluon propagator is described by two tensor structures, 

\begin{equation}
D^{ab}_{\mu\nu}(q)=\delta^{ab}\bigg(P^{T}_{\mu\nu} D_{T}(q_4,\vec{q})+P^{L}_{\mu\nu} D_{L}(q_4,\vec{q}) \bigg) \nonumber
\label{tens-struct}
\end{equation}
where the transverse and longitudinal projectors are defined by
\begin{equation}
P^{T}_{\mu\nu} = (1-\delta_{\mu 4})(1-\delta_{\nu 4})\left(\delta_{\mu \nu}-\frac{q_\mu q_\nu}{\vec{q}^2}\right) \quad , \quad
P^{L}_{\mu\nu} = \left(\delta_{\mu \nu}-\frac{q_\mu q_\nu}{{q}^2}\right) - P^{T}_{\mu\nu} \, ;
\label{long-proj}
\end{equation}
the transverse $D_T$ and longitudinal  $D_L$ propagators are given by
\begin{equation}
D_T(q)=\frac{1}{2V(N_c^2-1)}\left(\langle A_i^a(q) A_i^a(-q)\rangle-\frac{q_4^2}{\vec{q}^2} \langle A_4^a(q) A_4^a(-q)\rangle \right) \nonumber
\end{equation}

\begin{equation}
D_L(q)=\frac{1}{V(N_c^2-1)}\left(1+\frac{q_4^2}{\vec{q}^2} \langle A_4^a(q) A_4^a(-q)\rangle\right) \nonumber
\end{equation}

 In table \ref{tempsetup} we show the temperatures we have simulated up to now\footnote{Simulations in this section have been performed with the help of Chroma library \cite{chroma}; the FFT transforms have been done with the PFFT library \cite{pfft}.}, for a fixed physical spatial volume  $\sim(6.5\mbox{fm})^3$. For the determination of the lattice spacing we fit the string tension data in \cite{bali92}, using the functional form used in \cite{neccosommer}, in order to have a function $a(\beta)$.  

\begin{table}
\begin{center}
\begin{tabular}{cccccc}
\hline
Temp. (MeV) &	$\beta$ & $L_s$ &  $L_t$ & a [fm] & 1/a (GeV) \\
\hline
121 &	6.0000 & 64    & 	16 & 	0.1016 &  	1.9426 \\
162 &	6.0000 & 64    & 	12 & 	0.1016 & 	1.9426 \\
243 &	6.0000 & 64    & 	8 & 	0.1016 & 	1.9426 \\
260 &	6.0347 & 68    & 	8 & 	0.09502 & 	2.0767 \\
265 &	5.8876 & 52    & 	6 & 	0.1243 & 	1.5881 \\
275 &	6.0684 & 72    & 	8 & 	0.08974 & 	2.1989 \\
285 &	5.9266 & 56    & 	6 & 	0.1154 & 	1.7103 \\
290 &	6.1009 & 76    & 	8 & 	0.08502 & 	2.3211 \\
305 &	5.9640 & 60    & 	6 & 	0.1077	 &      1.8324 \\
305 &	6.1326 & 80    & 	8 & 	0.08077 & 	2.4432 \\
324 &	6.0000 & 64    & 	6 & 	0.1016	 &      1.9426 \\
486 &	6.0000 & 64    & 	4 & 	0.1016	 &      1.9426 \\
\hline
\end{tabular}
\end{center}
\label{tempsetup}
\caption{Lattice setup used for the computation of the gluon propagator at finite temperature.}
\end{table}

\begin{figure}[t] %  figure placement: here, top, bottom, or page
   \centering
   \subfigure{ \includegraphics[scale=0.25]{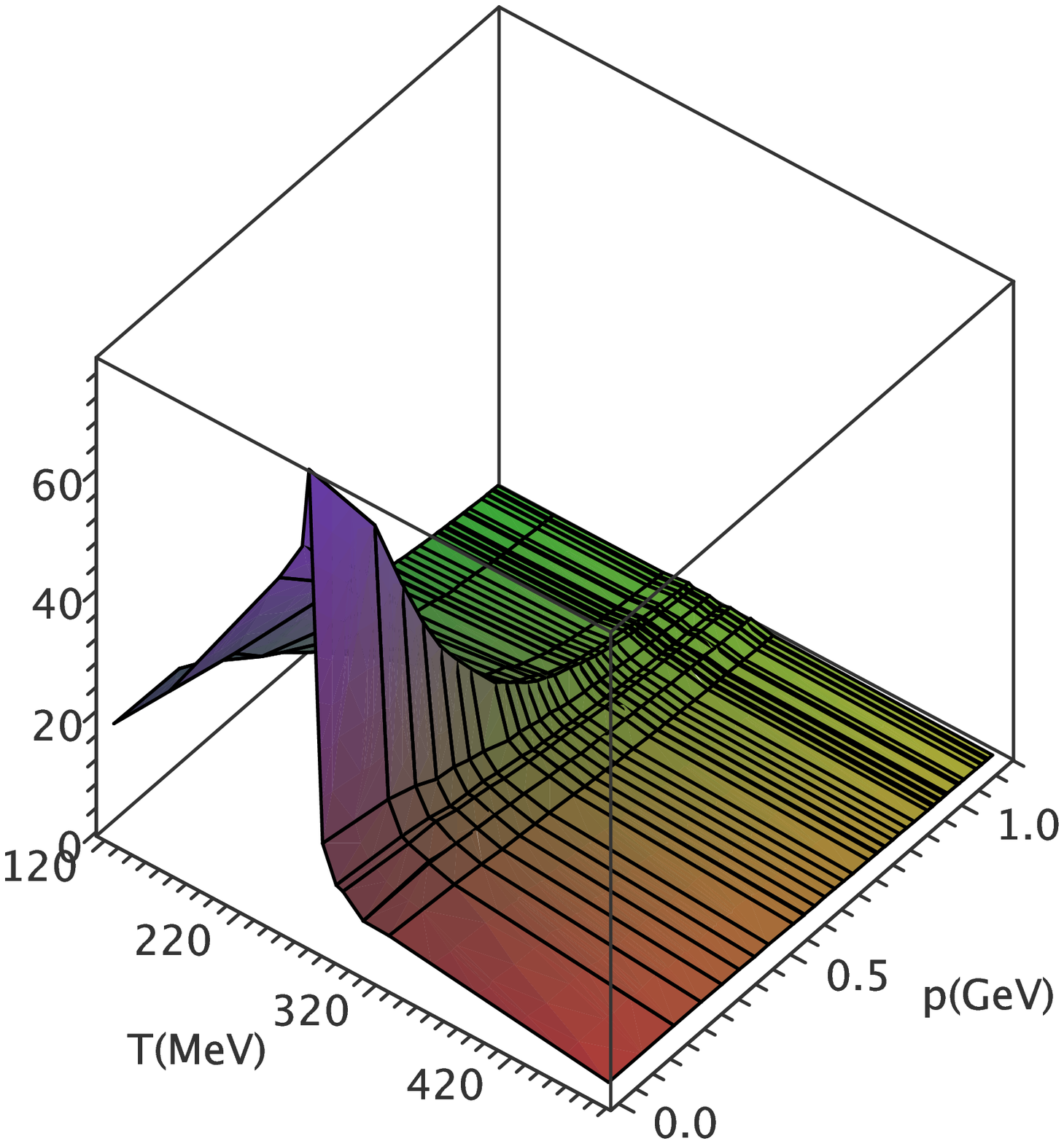} } \qquad
   \subfigure{ \includegraphics[scale=0.25]{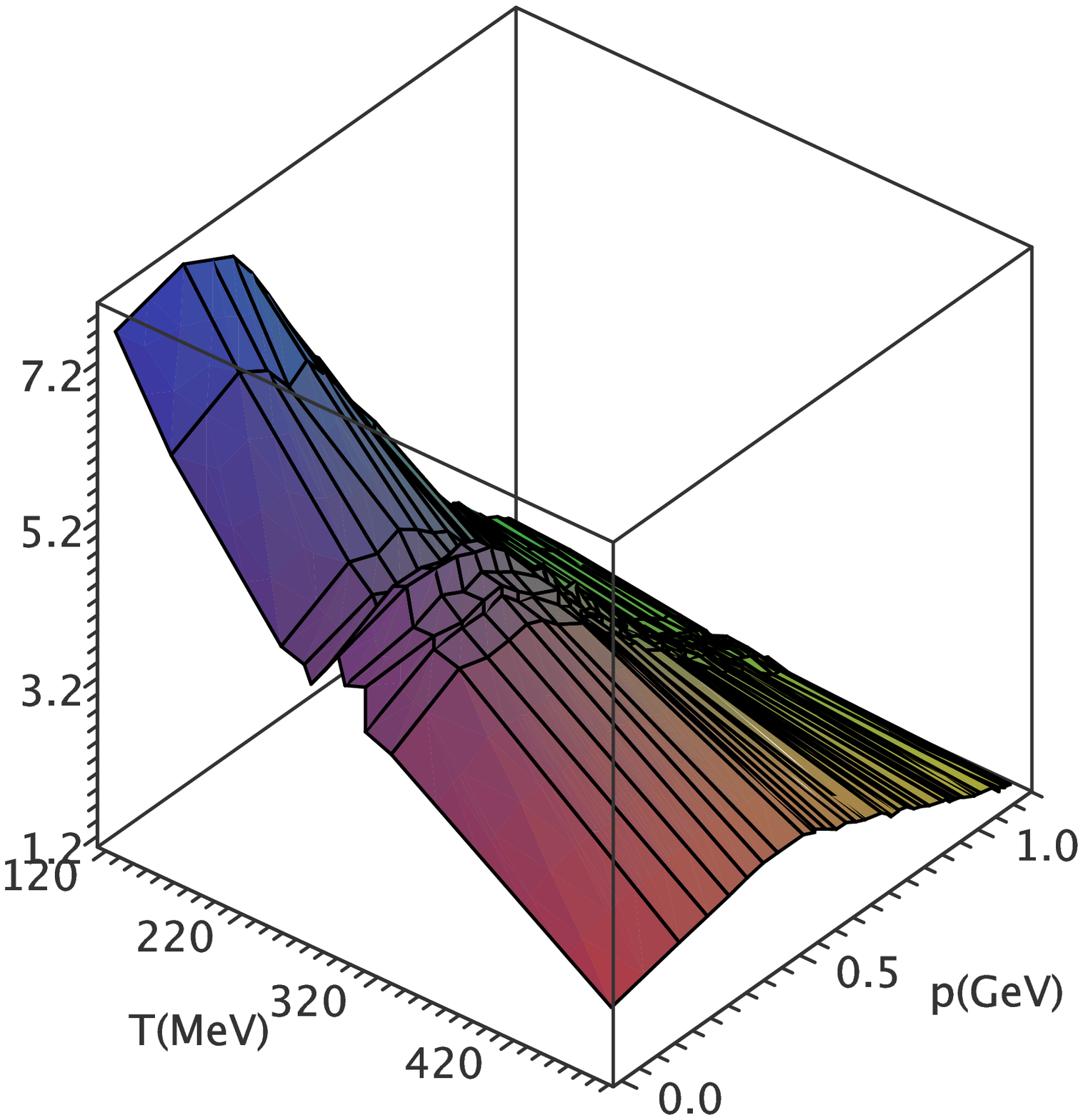} }
  \caption{Longitudinal (left) and transverse (right) gluon propagator as a function of momentum and temperature for a $\sim(6.5\mbox{fm})^3$ spatial lattice volume.}
   \label{fig:3dtemp}
\end{figure}

We resume the results obtained up to date in the 3d plots shown in figure \ref{fig:3dtemp}.  We see that the transverse propagator, in the infrared region, decreases with the temperature. Moreover, this component  exhibits a turnover for small momenta. The longitudinal component increases for temperatures below $T_c\sim 270\, \mbox{MeV}$. Then the data exhibits a discontinuity around $T_c$, and the propagator decreases for $T > T_c$. 

\section*{Acknowledgements}
We would like to thank the Berlin-Moscow-Adelaide group for sending us 
their data and for allowing to use it.
 
Paulo Silva acknowledges support 
by FCT under contract SFRH/BPD/40998/2007. Work supported by projects 
CERN/FP/123612/2011, CERN/FP/123620/2011 and PTDC/FIS/100968/2008, 
projects developed under initiative QREN financed by UE/FEDER through 
Programme COMPETE.


\begin{thebibliography}{99}

\bibitem{milc} This work was in part based on the MILC collaboration's public lattice gauge theory code. See http://physics.utah.edu/$\sim$detar/milc.html

\bibitem{ratios} O. Oliveira, P. J. Silva, Eur.Phys.J. \textbf{C62} (2009) 525, arXiv:0705.0964 [hep-lat].
\bibitem{bounds} O. Oliveira, P. J. Silva, Phys.Rev. \textbf{D79} (2009) 031501(R), arXiv:0809.0258 [hep-lat].
\bibitem{brasil} A. Cucchieri, T Mendes, PoS(LAT2007)297, arXiv:0710.0412 [hep-lat].
\bibitem{BMA09}
 I. L. Bogolubsky, E.-M. Ilgenfritz, M. M\"uller-Preussker, A. Sternbeck, Phys.Lett. \textbf{B676} (2009) 69, arXiv:0901.0736 [hep-lat].
\bibitem{OliSi12}
O. Oliveira, P. J. Silva, arXiv:1207.3029 [hep-lat].
\bibitem{lusobelga} D. Dudal, O. Oliveira, N. Vandersickel, Phys. Rev. \textbf{D81} (2010) 074505; A. Cucchieri, D. Dudal, T. Mendes, N. Vandersickel, arXiv:1111.2327 [hep-lat].
 \bibitem{chroma} R. G. Edwards, B. Jo\'{o}, Nucl. Phys. Proc. Suppl. \textbf{140} (2005) 832, arXiv:hep-lat/0409003.
\bibitem{pfft} M. Pippig, PFFT - An Extension of FFTW to Massively Parallel Architectures, Chemnitz University of Technology, 09107 Chemnitz, Germany Preprint 06.
\bibitem{bali92} G. Bali, K. Schilling, Phys. Rev. \textbf{D47} (1993) 661, arXiv:hep-lat/9208028.
\bibitem{neccosommer} S. Necco, R. Sommer, Nucl.Phys. \textbf{B622} (2002) 328-346, arXiv:hep-lat/0108008.

\end{thebibliography}
\end{document}